\documentclass[prl,twocolumn,showpacs,superscriptaddress]{revtex4}
\usepackage{graphicx}
\begin{document}
%\draft
\title{Analysis and interpretation of high transverse entanglement in
optical parametric down conversion}

\author{C. K. Law}
\affiliation{Department of Physics, The Chinese University of Hong
Kong, Shatin, Hong Kong, China}
\author{J. H. Eberly} \affiliation{Rochester Center for Quantum
Information, Rochester Theory Center for Optical Science and
Engineering, and Department of Physics and Astronomy, University
of Rochester, Rochester, NY 14627}

%%%%%%%%%%%%%%%%%%%%%%%%%%%%%%%%%%%%%%%%%%%%%%%%%%%%%%%%%%%%%
\begin{abstract}
Quantum entanglement associated with transverse wave vectors of
down conversion photons is investigated based on the Schmidt
decomposition method. We show that transverse entanglement
involves two variables: orbital angular momentum and transverse
frequency. We show that in the monochromatic limit high values of
entanglement are closely controlled by a single parameter
resulting from the competition between (transverse) momentum
conservation and longitudinal phase matching. We examine the
features of the Schmidt eigenmodes, and indicate how entanglement
can be enhanced by suitable mode selection methods.

\end{abstract}

%>>>> Include a list of keywords after the abstract

%\vspace{10mm} \pacs{PACS numbers: }

%%%%%%%%%%%%%%%%%%%%%%%%%%%%%%%%%%%%%%%%%%%%%%%%%%%%%%%%%%%%%

\date{\today}
\maketitle

Nonclassical correlation among photons has been one of the central
topics in quantum optics and quantum communication. In particular,
the property of quantum entanglement is now recognized as the key
to accomplishments exceeding the limit imposed by classical
physical laws. Entangled photons are experimentally available
through optical spontaneous parametric downconversion (PDC). Such
highly correlated photons have been employed in fundamental
experiments, e.g., to test the violation of Bell's inequality
\cite{bell1}, and to demonstrate quantum teleportation
\cite{tele1}. Significantly, down conversion photons provide
quantum entanglement in a rich diversity of forms beyond
polarization Bell states. For example, one has continuous
entanglement in field frequencies \cite{huang,Ian}, energy-time
variables \cite{energy1} and momenta \cite{momentum}. Altogether,
counting the discrete polarization and orbital angular momentum
variables \cite{angular1,angular2,angular3}, a full quantum
characterization of just two-photon states already requires
multiple parameters and the term `hyper-entanglement' has been
proposed in order to emphasize the complex structure of such
quantum states \cite{multi1}.

Here we focus on PDC entanglement that is associated with the
transverse wavevectors of signal and idler photon pairs. We are
especially interested in ways to reach states of anomalously high
entanglement. The PDC quantum state is sometimes called a biphoton
\cite{Klyshko}. Biphoton correlation is experimentally accessible
and several investigations have addressed this topic in regard to
the underlying orbital angular momentum structure
\cite{angular1,angular2,Torres} and applications in quantum
imaging \cite{imaging1} are being explored, including comparison
with two-beam correlation using spatially coherent classical light
\cite{boyd-etal}.  In this note we approach the subject using the
single-sum Schmidt decomposition technique \cite{knight}, which
yields for the first time a complete characterization of the
existing entanglement by determining the natural set of
bi-orthogonal mode pairs. This is in contrast to studies based on
familiar double-sum expansions in Gauss-Laguerre or Gauss-Hermite
modes.

To begin with, we consider the signal and idler photons of PDC
under the so-called type-II phase matching condition. The two
photons are distinguished by their orthogonal polarizations. The
explicit forms of biphoton states in three dimensional space are
complicated by the details of crystal properties and pump pulse
profiles \cite{rubin}. However, under plausible assumptions in the
paraxial approximation, the dependence of longitudinal components
of the wavevectors $k_z$ and $q_z$ may be eliminated in the
monochromatic limit \cite{angular2,monken}. Denoting ${\bf k}_
\bot$ and ${\bf q}_ \bot$ as transverse wavevectors of the signal
and idler photons, the biphoton state takes the approximate form:
$\left| \Psi \right\rangle = \int d{\bf k}_ \bot \int d{\bf q}_
\bot ~C \left( {{\bf k}_ \bot,{\bf q}_ \bot } \right)\left| {{\bf
k}}_ \bot, {{\bf q}_ \bot } \right\rangle.$ Here the biphoton
amplitude is given by \cite{angular2,monken},
\begin{equation}
\label{eq.Cdefn}
C \left( {{\bf k}_ \bot,{\bf q}_ \bot } \right) = {\cal E}_p
({{\bf k}_ \bot+{\bf q}_ \bot }) \Delta ({{\bf k}_ \bot-{\bf q}_\bot }),
\end{equation}
where ${\cal E}_p({\bf k}_\bot + {\bf q}_\bot)$ describes the
transverse profile of the pump field in wavevector space, and
$\Delta ({\bf k}_ \bot-{\bf q}_ \bot) $ is a purely geometrical
function that will be specified later. Note that if the amplitude
$C( {\bf k}_\bot,{\bf q}_\bot )$ were to separate into factors
depending only on ${\bf k}_\bot$ and ${\bf q}_\bot $, then the
state $|\Psi \rangle$ would be factorable (not entangled).

The Schmidt decomposition of $C( {\bf k}_\bot,{\bf q}_\bot )$
corresponds to the expansion,
\begin{equation}
\label{eq.Cexpansion}
C\left( {{\bf k}_ \bot ,{\bf q}_ \bot } \right) = \sum\limits_{n =
0}^\infty  {\sqrt {\lambda _n } } u_n ({\bf k}_ \bot) v_n ({\bf
q}_ \bot)
\end{equation}
where $u_n ({\bf k}_ \bot)$ and $v_n ({\bf q}_ \bot )$ are Schmidt
modes defined by eigenvectors of the reduced density matrices for
the signal and idler photons respectively, and $\lambda _n$ are
the corresponding eigenvalues \cite{knight}. Thus the mode
functions $u_n$ form a complete and orthonormal set, and the same
is true for the $v_n$. Because density matrices always have finite
trace, the Schmidt decomposition is always discrete, even when the
original specification of the state is naturally continuous, or
doubly continuous as in the present situation. This discreteness
is intrinsic, independent of any artificial box boundary
conditions.

The Schmidt decomposition (\ref{eq.Cexpansion}) provides the {\em
information eigenstates} of the two-particle system, which
characterizes the structure of entanglement by providing two
important pieces of information not available otherwise. First,
the Schmidt single-sum form of decomposition reveals exactly how
the photons are paired, i.e., if a signal photon is detected in
the mode $u_n$, then with certainty the idler photon must be in
the mode $v_n$. Second, the Schmidt eigenvalues $\lambda_n$ serve
to measure the degree of entanglement. This is usually discussed
in terms of the entanglement entropy $ E = - \sum\nolimits_n
{\lambda _n } \log _2 \lambda _n $. However, a more transparent
and experimentally more direct measure of entanglement is the
`average' number of Schmidt modes involved. The Schmidt number (or
participation ratio) $K$ provides this average: $K \equiv
1/\sum\nolimits_n {\lambda _n^2 }$. The larger the value of $K$,
the higher the entanglement \cite{BellK}. We remark that the
maximum value of $K$ is governed by the volume of accessible phase
space of the system under relevant physical constraints such as
energy-momentum conservation.

We first examine a class of biphoton amplitudes in which both
${\cal E}_p$ and $\Delta$ are represented by gaussian functions:
\begin{equation}
C_g({\bf k}_{\bot}, {\bf q}_{\bot} ) = {\cal N}_g e^{ {- |{\bf
k}_{\bot} +{\bf q}_{\bot}|^2 / \sigma_{\bot}^2}  }e^{-b^2 |{\bf
k}_{\bot}-{\bf q}_{\bot} |^2}.
\end{equation}
Here ${\cal N}_g$ is a normalization constant, and the parameters
$\sigma_{\bot}$ and $b^{-1}$ are the widths of the respective
gaussians, to which we give physical meaning below. With this
amplitude, it can be shown that the Schmidt eigenmodes are simply
the energy eigenfunctions of a two-dimensional isotropic harmonic
oscillator. The probability of finding a mode (in polar
coordinates) with radial and angular quantum numbers $n$ and $m$
is proportional to $\xi^{2n+|m|}$, where $\xi = ( {1 - {b\sigma _
\bot  } } )^2 /\left( {1 + b\sigma _ \bot  } \right)^2$. This
leads to a very convenient closed form for the Schmidt number:
\begin{equation} K = \frac{1}{4}\Big( b\sigma_\bot +
\frac{1}{b\sigma_\bot} \Big)^2. \label{eq.KGauss}
\end{equation}
Thus, $K$ increases as the ``control parameter" $b\sigma_ \bot$
increases or decreases, i.e., for both $b \sigma_ \bot \gg 1 $ and
$b \sigma_ \bot \ll 1$ (see Fig. 1). It is interesting that $K$
can be expressed in terms of the variance of the transverse
wavevector components, $K^{1/2}  = \left\langle {q_j^2 }
\right\rangle /\sqrt {\left\langle {s_{ + j}^2 } \right\rangle
\left\langle {s_{ - j}^2 } \right\rangle }$, where $s_{\pm j} =
(k_j \pm q_j)/\sqrt 2$ with the subscript $j$ referring to a
component on the transverse $xy$ plane. Such an expression
connects the abstract measure of entanglement with observable
fluctuations.

\begin{figure}
\includegraphics[width=6.6cm]{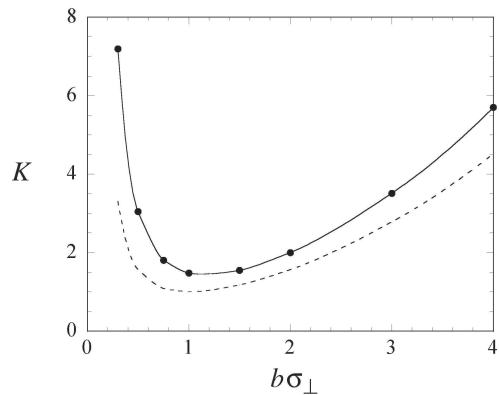}
\caption{Participation ratio $K$ as a function of $b
\sigma_{\bot}$ for the biphoton amplitude $C_g$ (dash line) and
$C_s$ (solid line). } \label{fig1}
\end{figure}

The biphoton amplitude $C_g$ can be recognized as a continuous
representation of two-mode squeezed states, and these states
maximize the EPR correlation for a given amount of entanglement
\cite{cirac}. Eq. (4) provides a useful estimation of degree of
entanglement according to the widths of the functions ${\cal
E}_p({\bf k}_\bot + {\bf q}_\bot)$ and ${\Delta}({\bf k}_\bot -
{\bf q}_\bot)$. In practice, $\Delta$ is determined by phase
matching in the longitudinal direction, and it is generally not a
gaussian. We now follow Monken, {\it et al}. \cite{monken} and
consider another class of biphoton amplitude:
\begin{equation}
\label{eq.CMonken} C_s({\bf k}_ \bot  ,{\bf q}_ \bot  ) = {\cal
N}_s e^{ {- |{\bf k}_{\bot} +{\bf q}_{\bot}|^2 / \sigma_{\bot}^2}
} {\rm sinc } \left[ { - b^2 | {{\bf k}_ \bot   - {\bf q}_ \bot  }
|^2 } \right].
\end{equation}
Here the transverse profile of the pump field ${\cal E}_p$ is
treated as a gaussian, and this gives physical meaning to the
width $\sigma_\bot$, and ${\cal N}_s$ is a normalization constant.
We have used the subscript $s$ in order to emphasize that $\Delta
({{\bf k}_ \bot-{\bf q}_ \bot })$ is now taken as the sinc
function arising from longitudinal momentum mismatch in the PDC
crystal. Spatial coherence is the physical effect that determines
$b$ via the relation $b^2= c L/4 \omega_p$, where $L$ is the
length of the crystal and $\omega_p$ is the pump frequency.

The amplitude (\ref{eq.CMonken}) is an approximation to actual
biphoton states realized in laboratories. This is because
technical details involving dispersion and birefringence effects
are omitted. However, it captures the main features imposed by
conservation of energy and momentum, which are the key constraints
on the state in the two-photon Hilbert space. To more fully
appreciate (\ref{eq.CMonken}) in terms of conservation rules, one
can see first that the gaussian with the argument ${\bf k}_
\bot+{\bf q}_ \bot$ is merely a statement of the uncertainty in
transverse momentum conservation, and second that the sinc
function's argument expresses energy conservation
$\omega_s=\omega_i= \omega_p/2$ in practice \cite{monken}.

In several previous studies of biphoton states, the sinc function
was approximated by unity in the small $L$ limit (or equivalently
the Fourier transform of $\Delta$ was approximated by a delta
function) \cite{angular2,Torres}. However, such an approximation
is inapplicable to our Schmidt analysis here because the
corresponding $C_s({\bf k}_ \bot ,{\bf q}_ \bot )$ would be
unbounded on the ${\bf k}_ \bot = -{\bf q}_ \bot$ manifold. Since
the sinc function ultimately limits the accessible transverse wave
vectors, it has to be fully accounted for in the equation. We also
note that exact analogs of the sum and difference arguments in
(\ref{eq.CMonken}) appear in wave functions of entangled states of
two massive particles when expressed in terms of center of mass
and relative coordinates \cite{Fedorov-etal}.

To carry out the complete Schmidt decomposition of
(\ref{eq.CMonken}), we make use of polar coordinates, writing:
${\bf k}_\bot = (k_\bot \cos \theta_k, ~ k_\bot\sin \theta_k)$ and
${\bf q}_\bot = (q_\bot \cos \theta_q, ~ q_\bot \sin \theta_q)$.
The biphoton amplitude can be decomposed in the form: $C_s({\bf
k}_ \bot  ,{\bf q}_ \bot  )= \sum\nolimits_{m } {{{e^{im(\theta _k
- \theta _q )}}} } \sqrt{P_m}F_m (k_ \bot ,q_ \bot  )/2 \pi,$
where $F_m(k_\bot,q_\bot)$ is a function of magnitude $k_\bot$ and
$q_\bot$, and is normalized according to $\int_0^\infty
{\int_0^\infty  {} |F_m (k_ \bot ,q_ \bot  )|^2k_ \bot  q_ \bot }
dk_ \bot  dq_ \bot =1$. We identify the integer $m$ as a quantum
number of orbital angular momentum. Therefore $P_m$ is the
probability of finding the two photons with opposite orbital
angular momentum numbers $m$ and $-m$.

Further decomposition of $F_m (k_ \bot ,q_ \bot) \sqrt {k_\bot
q_\bot}$ is a simpler task due to its lower dimensionality as
compared with  $C_s({\bf k}_ \bot  ,{\bf q}_ \bot  )$. According
to the Schmidt decomposition scheme, we have: $F_m (k_\bot ,q_
\bot )\sqrt {k_ \bot  q_ \bot }  = \sum\nolimits_{n} {\sqrt
{\gamma _{n,m} } } \phi _{n,m} (k_ \bot  )\phi _{n,m} (q_ \bot )$,
where the $\phi_{n,m}$ are the same for idler and signal photons
because of the symmetry assumed in Eq. (\ref{eq.CMonken}).
Consequently the Schmidt decomposition of the biphoton amplitude
reads:
\begin{equation}
\label{eq.CmnSum} C_s({\bf k}_ \bot  ,{\bf q}_ \bot  ) =
    \sum\limits_{m =  - \infty}^\infty \sum\limits_{n=0}^{\infty}
    \sqrt{ \lambda_{nm}} u_{n,m}
({\bf k}_\bot) u_{n,-m} ({\bf q}_\bot)
\end{equation}
where $\lambda_{nm} \equiv P_m \gamma _{nm}$ and $u_{n,m} ({\bf
k}_\bot) \equiv e^{im\theta _k} \phi _{n,m} (k_ \bot  ) / \sqrt{2
\pi k_ \bot}$ are normalized Schmidt mode functions in wavevector
space. Eq. (6) reveals that quantum entanglement involves orbital
angular momenta and the magnitude of transverse wavevectors.

\begin{figure}
\includegraphics[width=6.6cm]{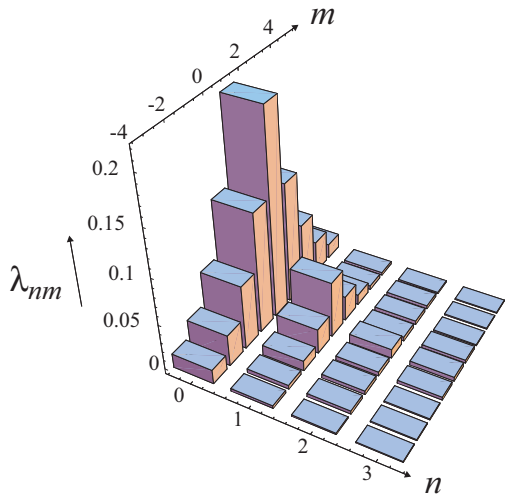}
\caption{Distribution of $\lambda_{nm}$ of the biphoton amplitude
$C_s$ with $b \sigma_{\bot}=0.25$. } \label{fig2}
\end{figure}

The gaussian and the sinc function in (\ref{eq.CMonken}) compete
with each other by enforcing correlations in opposite directions.
High entanglement is achieved only when one of them becomes
dominant, i.e., $\sigma_{\bot}b \gg 1$ or $ \sigma_{\bot}b \ll 1$.
We have numerically performed the Schmidt decomposition for
various values of $\sigma_{\bot}b$. In Fig. 1, we show the Schmidt
number $K = 1/\sum\nolimits_{nm}^{} \lambda _{nm}^2$ as a function
$\sigma_{\bot} b$. Consistent with the description above, we see
that either a decrease or increase in $\sigma_{\bot} b$ can lead
to an increased degree of entanglement. For example, $K\approx
7.2$ when $\sigma_{\bot} b =0.3$. The minimum Schmidt number $K
\approx 1.4$ occurs at $\sigma_{\bot} b \approx 1$. The figure
also indicates that $C_s$ is more entangled than $C_g$ at the same
parameter value $\sigma_{\bot} b$.

\begin{figure}
\includegraphics[width=8.6cm]{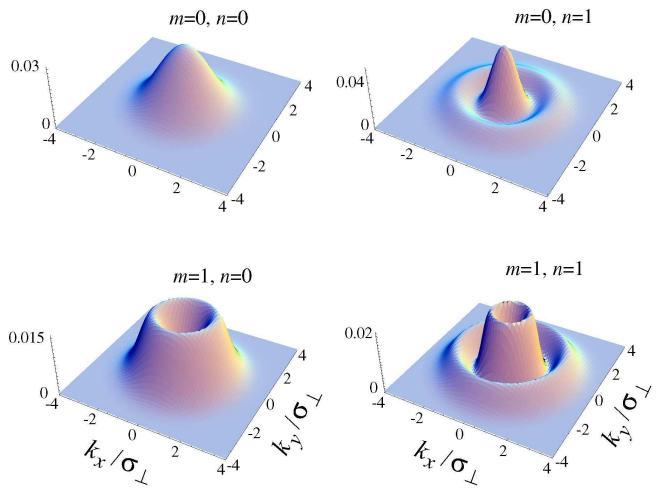}
\caption{An illustration of first few Schmidt eigenmodes $|u_{n,m}
({\bf k}_\bot)|^2$ of the biphoton amplitude $C_s$ with $b
\sigma_{\bot}=0.25$.} \label{fig3}
\end{figure}

To learn how various modes contribute to the entanglement, we plot
in Fig. 2 the distribution of $\lambda_{nm}^2$ for the case
$b\sigma_{\bot}=0.25$. Such a control parameter value corresponds
to a moderately high entanglement regime where conservation of
transverse momentum dominates. Small values of $b\sigma_{\bot}$
are typically accessible in experiments using very thin crystals.
In Fig. 2 we clearly see that most probabilities are found on the
$n=0$ manifold, where the quantum number $n$ equals the number of
nodes in the radial direction (Fig. 3). Therefore quantum
entanglement is mainly manifest in the angular correlations. In
fact, the nature of quantum entanglement can be appreciated from
the probability distribution associated with the photon amplitude
(Fig. 4). Along the $k=q$ manifold, we see a narrow peak emerge
near $\Delta \theta = \theta_k -\theta_q = \pi$. For smaller
values of $b\sigma_{\bot}$ (i.e., higher entanglement), the more
pronounced is the peak observed. In other words ${\bf k}_ \bot$ is
almost locked near $- {\bf q}_ \bot$ for sufficiently large
transverse wavevectors. This demonstrates an interesting
connection between quantum entanglement and localization in the
transverse momentum space. In fact, recent studies have begun to
examine the role of entanglement in localization problems
\cite{Fedorov-etal,rau}.

Fig. 4 suggests that wave vectors with large transverse magnitude
are `more entangled'. Therefore higher entanglement may be
achieved by simply choosing wavevectors with sufficiently large
transverse components. This may be realized by adopting certain
filtering procedures. As a demonstration, we consider an amplitude
\begin{equation}
{\cal A}({\bf k}_{\bot}, {\bf q}_{\bot} )= C_s({\bf k}_{\bot},
{\bf q}_{\bot} ) \theta(k_{\bot}-\mu_c)\theta(q_{\bot}-\mu_c)
\end{equation}
in which low transverse wavevectors smaller than $\mu_c$ are
removed by the unit step functions $\theta(x)$. We have performed
a Schmidt decomposition of ${\cal A}({\bf k}_{\bot}, {\bf
q}_{\bot} )$ numerically. For the case $b\sigma_{\bot}=0.25$ and
$\mu_c=2\sigma_{\bot}$, the corresponding Schmidt number is found
to be $K \approx 17.2$. This is substantially higher than the
original value $K=10.2$. The enhancement is more drastic for the
almost disentangled case $b\sigma_{\bot}=1$. In this case we find
$K \approx 26$ by using $\mu_c=\sigma_{\bot}$. Such an enhancement
via amplitude filtering exploits the long tail of the sinc
function, and it has a cost in terms of detection rate. Physically
it means that there is uncertainty of projecting the two photon
amplitude onto ${\cal A}({\bf k}_{\bot}, {\bf q}_{\bot} )$. For
the $b\sigma_{\bot}=1$ example above, the probability of finding
the system in ${\cal A}({\bf k}_{\bot}, {\bf q}_{\bot} )$ is about
$6\%$.

\begin{figure}
\includegraphics[width=6.6cm]{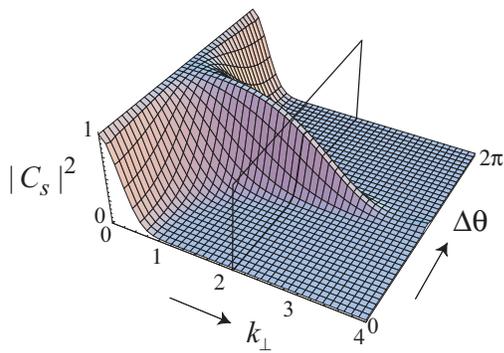}
\caption{A plot of the two photon amplitude square $|C_s({\bf
k}_{\bot}, {\bf q}_{\bot} )|^2 $ in the $k_{\bot} = q_{\bot}$
subspace. For convenience, the peak value is normalized to one. A
cutting plane is shown for the entanglement enhancement scheme
discussed in the text. The parameter is $b \sigma_{\bot}=0.25$. }
\label{fig4}
\end{figure}

To summarize, we described a method to analyze and interpret
physically realistic cases of high transverse entanglement in
parametric down conversion, based on the Schmidt decomposition
method. For the two classes of two-photon amplitudes defined by
$C_g$ and $C_s$, the degree of entanglement is controlled by a
single parameter $b\sigma_{\bot}$. We determined the Schmidt
number as a function of $b\sigma_{\bot}$ and identified two
experimentally accessible regimes $b\sigma_{\bot} \gg 1$ and
$b\sigma_{\bot} \ll 1$ in which values of entanglement higher than
reported to date can be studied. In addition we proposed a method
to enhance entanglement via large transverse wave vector
selection. In fact one sees many possibilities to engineer
entanglement through mode selection and manipulation of the shape
of the pump pulse profile \cite{Torres}. The Schmidt decomposition
presented here provides a basic framework for investigation of
such possibilities.

\bigskip\noindent
Acknowledgments: We acknowledge conversations with colleagues,
including K. Banaszek, K.W. Chan, M.V. Fedorov, A. Gatti, J. Howell,
M. Ivanov, A. Sergienko and I.A. Walmsley, and support from grants
NSF PHY00-72359 and ARO DAAD19-99-1-0215, and RGC Project No. 401603.


\begin{thebibliography}{1}

\bibitem{bell1} Z.Y. Ou and L. Mandel, Phys. Rev. Lett. {\bf 61},
50 (1988); Y.H. Shih and C.O. Alley, Phys. Rev. Lett. {\bf 61},
2921 (1988).

\bibitem{tele1}D. Bouwmeester {\it et al.},  Nature {\bf 390}, 575
(1997); D. Boschi {\it et al.}, Phys. Rev. Lett. {\bf 80}, 1121
(1998).

\bibitem{huang} H. Huang and J.H. Eberly, J. Mod. Opt.
{\bf 40}, 915 (1993).

\bibitem{Ian} C.K. Law, I.A. Walmsley, and J.H. Eberly,
Phys. Rev. Lett. {\bf 84}, 5304 (2000); W.P. Grice, A.B. U'Ren,
and I.A. Walmsley, Phys. Rev. A {\bf 64}, 063815 (2001).

\bibitem{energy1} P.G. Kwiat, A. M.
Steinberg, and R.Y. Chiao, Phys. Rev. A {\bf 47}, R2472 (1993); J.
Brendel {\it et al.}, Phys. Rev. Lett. {\bf 82}, 2594 (1999).

\bibitem{momentum} J.G. Rarity and P.R. Tapster,
Phys. Rev. Lett. {\bf 64}, 2495 (1990).

\bibitem{angular1} A. Mair {\it et al.},
Nature {\bf 412}, 313 (2001).

\bibitem{angular2} S. Franke-Arnold {\it et al.}, Phys. Rev. A
{\bf 65}, 033823 (2002).

\bibitem{angular3} H.H. Arnaut and G.A. Barbosa,
Phys. Rev. Lett. {\bf 86}, 5209 (2001); G.A. Barbosa and H.H.
Arnaut, Phys. Rev. A {\bf 65}, 053801 (2002).

\bibitem{multi1} P.G. Kwiat, J. Mod.
Opt. {\bf 44}, 2173 (1997); M. Atature {\it et al.}, Phys. Rev. A
{\bf 65}, 023808 (2002).

\bibitem{Klyshko} D.N. Klyshko, JETP Lett. {\bf 6}, 23 (1967).


\bibitem{Torres} J.P. Torres {\it et al.}, Phys. Rev. A {\bf 67},
052313 (2003).

\bibitem{imaging1} A. Gatti, E. Brambilla, and L.A. Lugiato,
Phys. Rev. Lett. {\bf 90}, 133603 (2003); A.F. Abouraddy {\it et
al.}, Phys. Rev. Lett. {\bf 87}, 123602 (2001).

\bibitem{boyd-etal} R.S. Bennink, S.J. Bentley, and R.W.
Boyd, Phys. Rev. Lett. {\bf 89}, 113601 (2002).

\bibitem{knight} See A. Peres, {\em Quantum Theory: Concepts and
Methods} (Kluwer Academic, 1995), and A. Ekert and P.L. Knight,
Am. J. Phys. {\bf 63}, 415 (1995).

\bibitem{rubin} M.H. Rubin,
Phys. Rev. A {\bf 54}, 5349 (1996).

\bibitem{monken} C.H. Monken, P.H. Souto Ribeiro, and S. Padua,
Phys. Rev. A {\bf 57}, 3123 (1998); S.P. Walborn, A.N. de
Oliveira, and C.H. Monken, Phys. Rev. Lett. {\bf 90}, 143601
(2003).

\bibitem{BellK}  See R. Grobe, K. Rz\c{a}\.zewski and J.H. Eberly,
J. Phys. B {\bf 27}, L503 (1994). A disentangled (product) state
corresponds to $K=1$, i.e., there is only one term in the Schmidt
decomposition.  Bell states of two-photon polarization have $K=2$.
Only two-particle states in continuous Hilbert spaces can comprise
a large number of Schmidt modes and have very large $K$ values.

\bibitem{cirac} G. Giedke {\it et al.}, Phys. Rev. Lett. {\bf 91},
107901 (2003).

\bibitem{Fedorov-etal} We have recently analyzed entanglement
arising in two-particle breakup of massive particles, as in
nuclear fission or molecular dissociation.  See M.V. Fedorov {\it
et al.}, to be submitted, 2003.

\bibitem{rau} A connection between entanglement and
localization of two particles was recently discussed by A.V. Rau,
J.A. Dunningham, and K. Burnett, Science {\bf 301}, 1081 (2003).

\end{thebibliography}
\end{document}